\documentclass[10pt, twocolumn]{IEEEtran}
\usepackage{graphicx,amsmath,amssymb,cite}
\usepackage{subfigure}
\usepackage{graphicx,subfigure,amsmath,amssymb,amsfonts,cite,float,color,longtable,rotating,diagbox,booktabs,array}
\usepackage{float,color}
\ifCLASSINFOpdf
\else
\fi

\begin{document}

\title{Artificial-noise-aided Secure Multicast Precoding for Directional Modulation Systems}

\author{Feng~Shu, Ling~Xu, Jiangzhou~Wang,~Wei Zhu, and~Zhou Xiaobo
\thanks{This work was supported in part by the National Natural Science Foundation of China (Nos. 61472190, and 61501238).}
\thanks{Feng~Shu,~Ling~Xu,{\color{blue}{~Wei Zhu, and~Zhou Xiaobo}} are with School of Electronic and Optical Engineering, Nanjing University of Science and Technology, 210094, CHINA. E-mail:\{shufeng, xuling\}@njust.edu.cn}
\thanks{Jiangzhou~Wang is  with  the School of Engineering and Digital Arts, University of Kent, Canterbury Kent CT2 7NT, United Kingdom}
}

\maketitle
\begin{abstract}
In multi-cast scenario, all desired users are divided into $K$ groups. Each group receives its own individual confidential message stream. Eavesdropper group aims to intercept $K$ confidential-message streams. To achieve a secure transmission, two secure schemes  are proposed: maximum group receive power plus null-space (NS) projection (Max-GRP plus NSP) and leakage.  The former obtains its precoding vector per group by maximizing its own group receive power  subject to the orthogonal constraint, and  its AN projection matrix  consist of all bases of NS of all desired steering vectors from all groups. The latter attains its desired precoding vector per group by driving the current confidential message power to its group steering space and reducing its power leakage to eavesdropper group and other $K-1$ desired ones by maximizing signal to leakage and noise ratio (Max-SLNR). And its AN projection matrix is designed by forcing AN power into the eavesdropper steering space by viewing  AN as a useful signal for eavesdropper group and maximizing AN to leakage-and-noise ratio (Max-ANLNR). Simulation results show that the proposed two methods are better than conventional method in terms of both bit-error-rate (BER) and secrecy sum-rate per group. Also, the leakage scheme  performs better than Max-GRP-NSP , especially in the presence of direction measurement errors. However, the latter requires no channel statistical parameters and thus is simpler compared to the former.
\end{abstract}


%
\IEEEpeerreviewmaketitle

\section{Introduction}
{In the recent several years,  physical-layer security in wireless networks has attracted more and more research activities from both academia and industry \cite{Wyner1975,zhao,YAN1,zou1,Li1,Li2}. In \cite{Wyner1975}, the author's pioneer research work has laid a foundation for physical-layer security.}  As a physical layer secure transmit technique in line-of-propagation (LoP) channel, directional modulation (DM) has made rapid progress in many aspects by using antenna array with the help of aided artificial noise (AN) \cite{maha,kalantari,babakhani,daly,Ding,Hu1,Wu}. To enhance security,  the symbol-level precoder in \cite{kalantari} was presented by using the concept of constructive interference in directional modulation with the goal of reducing the energy consumption at transmitter. In the presence of  direction measurement error,  the authors in \cite{Hu1,Wu} proposed two new robust DM synthesis methods for two different application scenarios: single-desired user and multi-user broadcasting by fully exploiting the statistical properties of direction measurement error. Compared to existing non-robust methods, the proposed robust methods can achieve at least an order-of-magnitude bit error rate (BER) performance improvement along desired directions. In general, the DM in LoP channel might harvest a high performance gain along the desired directions via confidential-message beamforming and  degrades the performance of eavesdroppers at undesired directions with the help of AN projection operation. Finally,  the goal of secure transmission is realized.

\cite{maha, kari1} established a unified framework for physical layer multi-casting to multiple co-channel groups, where multiple independent data streams are transmitted to groups of users by the multiple antennas. If one group of eavesdropper appears, then how to achieve a secure transmission in such a situation is an interesting and important research topic. This secure problem includes twofold:  the privacy protection among desired groups and the leakage of all confidential messages from all desired groups to the group of eavesdroppers. We will address this topic from the standpoint of physical layer security by using DM  in this paper, where the LoP channel is considered and  two precoding and AN projecting schemes for this multi-cast communications will be proposed.

The remainder are organized as follows: Section II describes the system model in multi-cast scenario. Two AN-aided precoding schemes are proposed in Section III.  Section IV presents the simulation results. Finally, we make conclusions in Section V.

Notations: throughout the paper, matrices, vectors, and scalars are denoted by letters of bold upper case, bold lower case, and lower case, respectively. $(\cdot)^T$, $(\cdot)^*$， and $(\cdot)^H$ denote transpose,~conjugate,~and conjugate transpose,~respectively.  Matrices $\textbf{I}_N$ denotes the $N\times N$ identity matrix and $\textbf{0}_{M\times N}$ denotes $M\times N$ matrix of all zeros. $\text{tr}(\cdot)$ denotes matrix trace.

\section{System Model}
\begin{figure}[tp]
  \centering
  \includegraphics[width=7.5cm]{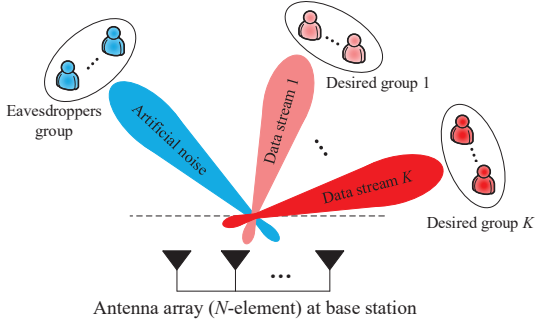}\\
  \caption{Schematic diagram of multicast DM  system}
  \label{Multicast DM}
\end{figure}

Fig.~\ref{Multicast DM} plots the schematic diagram of a multi-cast multiuser directional modulation system in LoP channel with perfect CSI konwledge. In this system, there is one base station (BS), $K$ groups of desired users, and one group of eavesdroppers. BS employs an $N$-antenna array. Desired group $k$ is composed of $T_k$ desired users. Eavesdropper group consists of $M$ eavesdroppers. {\color{blue}{All users are equipped with a single antenna.}} In Fig.~\ref{Multicast DM},  antenna array at BS broadcasts $K$ independent streams of confidential messages to $K$ different desired groups, respectively and securely so that all users in the eavesdropper group can't intercept any one of $K$ streams of confidential messages. In Fig.~\ref{Multicast DM}, {\color{blue}{ the normalized steering vector for the $i$th user of the $k$th desired group with direction angle $\theta_{d,ki}$ is
$\mathbf{h}(\theta_{d,ki})=1/\sqrt{N}[e^{j2\pi\psi_{\theta_{d,ki}}(1)}, \cdots, e^{j2\pi\psi_{\theta_{d,ki}}(N)}]^T$, where $\psi_{\theta_{d,ki}}(n)=(n-(N+1)/2)d\cos\theta_{d,ki}/\lambda$ with $d$ being antenna spacing and $\lambda$ being the wavelength of transmit carrier.
}}{\color{blue}{Therefore,}} the total steering channel matrix for $k$th desired group can be defined as
\begin{align}\label{equation 4}
\mathbf{H}(\theta_{dk})=[\mathbf{h}(\theta_{dk,1}), \mathbf{h}(\theta_{dk,2}), \cdots, \mathbf{h}(\theta_{dk,T_k})],
\end{align}
where $T_k$ denotes the number of desired users in  $k$th group. In (1), $\mathbf{H}(\theta_{dk})$ is an $N\times T_k$ matrix. Similarly, the $N\times M$ total steering channel matrix of the eavesdropper group is given by
$\mathbf{H}(\theta_e)=[\mathbf{h}(\theta_{e,1}),~\cdots,~\mathbf{h}(\theta_{e,M})]$. The transmit baseband signal is written as
\begin{align}\label{equ s}
\setlength{\abovedisplayskip}{1pt}
\setlength{\belowdisplayskip}{1pt}
\mathbf{s}=\alpha_1\beta_1\sqrt{P_s}\sum_{k=1}^K\mathbf{v}_kx_k+\alpha_2\beta_2\sqrt{P_s}\mathbf{T}_{AN}\mathbf{z}
\end{align}
where $x_k$ is the confidential message transmitted to the $k$th desired group, $\mathbf{v}_k$ is the corresponding beamforming vector of group $k$  for  confidential message $x_k$,  $P_s$ is the total transmit power,~$\mathbf{T}_{AN}$ is the AN projection matrix, and $\mathbf{z}$ is the AN vector, satisfying $\mathbf{z}\sim\mathcal{C}\mathcal{N}(0,\mathbf{I}_{N-\sum_{k=1}^KT_k})$. In (\ref{equ s}), $\beta_1$ and $\beta_2$ are the power allocation factors of confidential message and AN, with the constraint $ \beta_1^2+\beta_2^2=1$. A large value of $\beta_1$ means that more power is allocated to transmit confidential messages and less power is used as secure protection. How to choose the optimal values of $\beta_1$ and $\beta_2$ is a hard problem and depends on application scenarios. Parameter $\alpha_1$  is the normalized power factor of confidential message satisfying $\alpha_1^2\mathbf{E}\{\sum_{k=1}^K\sum_{i=1}^K\mathbf{v}_i^H\mathbf{v}_kx_kx_i^H\}=1$.
To simplify the above expression, we have $\alpha_1=1/\sqrt{\gamma K}$, where $\gamma$ is the normalized factor of signal constellation of digital modulation. For example, when quadrature phase shift keying (QPSK) is adopted, $\gamma=1/\sqrt{2}$, then
$\alpha_1=1/\sqrt{2K}$. The normalized AN power factor $\alpha_2$ ensures $\alpha_2\mathbf{E}\{\text{tr}[\mathbf{T}_{AN}\mathbf{z}\mathbf{z^H}\mathbf{T}_{AN}^H]\}=1$,
which can be simplified as $\alpha_2=1/\sqrt{\sigma_z^2tr(\mathbf{T_{AN}}\mathbf{T}_{AN}^H)}$. When all $N-\sum_{k=1}^{K}T_k$ columns of $\mathbf{T}_{AN}$ are orthogonal with each other and normalized, $\text{tr}(\mathbf{T}_{AN}\mathbf{T}_{AN}^H)=N-\sum_{k=1}^{K}T_k$. The received signal vector of desired group $k$ is
\begin{align}\label{Desire-Rx-Sig}
&\mathbf{y}(\theta_{dk})=\alpha_1\beta_1\sqrt{P_s}\mathbf{H}^H(\theta_{dk})\mathbf{v}_kx_k+\alpha_1\beta_1\sqrt{P_s}\mathbf{H}^H(\theta_{dk})\nonumber\\
&\sum_{i=1,i\neq k}^K\mathbf{v}_ix_i
+\alpha_2\beta_2\sqrt{P_s}\mathbf{H}^H(\theta_{dk})\mathbf{T}_{AN}\mathbf{z}+\mathbf{n}_{dk}
\end{align}
where  $n_{dk}\sim\mathcal{C}\mathcal{N}(0, \sigma_d^2\mathbf{I}_{T_k})$.
Similarly, the received signal vector of the eavesdropper group can be expressed as
\begin{align}\label{Eav-Rx-Sig}
&\mathbf{y}(\theta_e)=\alpha_1\beta_1\sqrt{P_s}\mathbf{H}^H(\theta_e)\sum_{k=1}^K\mathbf{v}_kx_k
+\alpha_2\beta_2\sqrt{P_s}\mathbf{H}^H(\theta_e)\bullet\nonumber\\&~~~~~~\mathbf{T}_{AN}\mathbf{z}+\mathbf{n}_e,
\end{align}
where $\mathbf{n}_e\sim\mathcal{C}\mathcal{N}(0, \sigma_e^2\mathbf{I}_M)$.
\begin{table*}[ht]
\caption{Complexity comparison(FLOPs).}
\newcommand{\tabincell}[2]{\begin{tabular}{@{}#1@{}}#2\end{tabular}}
\centering
\begin{footnotesize}
\begin{tabular}{|l|c|c|c}
\hline
Methods &Complexity as a function of $K$  & Complexity as a function of T
\\
\hline
Max-GRP-NSP
&\tabincell{l}{$O((7T^2N+3T^3)K^2$$+(-12T^2-4TN^2-3T^3-NT)K$\\$+(7T^2N+7TN^2+2N^3+T^3+N^2+NT))$}
&\tabincell{l}{$O((3K^2-3K+1)T^3$$+(7K^2N-12KN+7N)T^2$\\$+(7N^2-4KN^2+N-NK)T+(N^2+2N^3))$}\\
\hline
\tabincell{l}{Max-SLNR plus \\Max-ANLNR} &$O(2TN^2K+(3MN^2+TN^2+4N^3))$ &$O((2KN^2+N^2)T+(3MN^2+4N^3))$\\
\hline
BD
&\tabincell{l}{$O(3T^2NK^2$$+(T^3-T^2N+MNT+TN$\\$-M^2T-2N^2T)K$\\$+(-T^3+2M^2N-MNT)$$+(+T^2M-TN+MN+MN^2+N^3))$} &\tabincell{l}{$O((K-1)T^3$$+(3K^2N-KN+M)T^2+$\\$(MNK-MN+KN-N-M^2K-2N^2K)T$\\$+(2M^2N+MN+MN^2+N^3))$}\\
\hline
\end{tabular}
\end{footnotesize}
\end{table*}
\section{Two efficient schemes of precoding and AN projection}
In this section, we will present two methods to design the confidential message precoding vector $\mathbf{v}_k$ and AN projection matrix $\mathbf{T}_{AN}$ in (2). The first scheme devises the confidential message precoder of maximizing group receive power (Max-GRP) subject to the orthogonal constraint and the AN projection matrix based on null-space projection (NSP) rule. The second scheme uses the leakage concept to design both precoding vectors and AN projection matrix.  Due to the use of the statistical property of AN and channel noise, the latter provides a better and more robust performance than the former, which will be confirmed in our simulation section.
\subsection{Proposed Max-GRP plus NSP scheme}
Firstly, let us define the complement of the steering matrix of  group $k$ as ${\bf{H}}_{d,-k}=\big[{\bf{H}}\left(\theta _{d1}\right)$,$~\cdots,~{\bf{H}}\left(\theta _{d(k-1)}\right),~{\bf{H}}\left(\theta _{d(k+1)}\right),~\cdots,~{\bf{H}}\left(\theta _{dK}\right)\big]$. Then, the optimization problem of maximizing group receive power (Max-GRP) of group $k$ is casted as
\begin{align}\label{Max-RGP 1}
&\max\limits_{{{\bf{v}}_k}}{\kern 1pt}~~~~~~{\bf{v}}_k^H{{\bf{H}}({\theta_{dk}})}{\bf{H}}^H({\theta_{dk}}){{\bf{v}}_k} \nonumber\\
&\text{subject~to}~~{\bf{H}}_{d,-k}^H {\bf{v}}_k = 0.
\end{align}
The objective function in (5) denotes the receive power summation for all users of group $k$. The constraint ${\bf{H}}_{d,-k}^H {\bf{v}}_k = 0$ means that the confidential message of desired group $k$ will be transmitted through the corresponding null spaces of all the remaining desired groups by the optimization variable $\mathbf{v}_k$. The optimization problem in (5) aims to enhance the receive quality of confidential message for desired group $k$ by maximizing the receive power sum of all users of group $k$.
To solve the above problem, channel matrix ${\bf{H}}_{d,-k}$ is first decomposed as the singular-value decomposition (SVD)
\begin{equation}\label{equation 22}
\begin{array}{c}
\begin{split}
&{\bf{H}}_{d,-k}=\\
 &\left[ {\begin{array}{*{20}{c}}
{{\bf{U}}_{d,-k}^{\left( 1 \right)}}&{{\bf{U}}_{d,-k}^{\left( 0 \right)}}
\end{array}} \right]\left( {\begin{array}{*{20}{c}}
{{\bf\Sigma}_{d,-k}^{\left( 1 \right)}}&{\bf{0}}\\
{\bf{0}}&{\bf{0}}
\end{array}} \right){\left[ {\begin{array}{*{20}{c}}
{{\bf{V}}_{d,-k}^{\left( 1 \right)}}&{{\bf{V}}_{d,-k}^{\left( 0 \right)}}
\end{array}} \right]^{\rm{H}}}
\end{split}
\end{array}
\end{equation}
where ${\bf\Sigma}_{d,-k}^{\left( 1 \right)}$ is a ${L_k} \times {L_k}$ diagonal matrix, and ${\bf{V}}_{d,-k}^{\left( 0 \right)}$  consists of the last $\left( {N - {L_k}} \right)$ right singular vectors corresponding to $N-{L_k}$ zero singular values. Define  ${{\bf{F}}_k}={\bf{V}}_{d,-k}^{\left( 0 \right)}$ , and ${{\bf{v}}_k} = {{\bf{F}}_k}{{\bf{u}}_k}$, then the optimization problem in (\ref{Max-RGP 1}) is converted into
\begin{equation}\label{equation 24}
\begin{array}{l}
\max\limits_{{{\bf{u}}_k}}~~~~~~~~~{\bf{u }}_k^H{\bf{F}}_k^H{{\bf{H}}(\theta_{dk})}{\bf{H}}^H(\theta_{dk}){{\bf{F}}_k}{{\bf{u}}_k}\\
\text{subject to}~~~{\bf{u }}_k^H{{\bf{u }}_k} = 1,
\end{array}
\end{equation}
which means that ${{\bf{u }}_k}$ is the eigenvector corresponding to the largest eigenvalue of matrix ${\bf{F}}_k^H{{\bf{H}}(\theta_{dk})}{\bf{H}}^H(\theta_{dk}){{\bf{F}}_k}$. The design of ${{\bf{v}}_k}$ has been completed. In the following, we
 design the AN projection matrix ${\bf{T}}_{AN}$. The basic principle is to eliminate the influence of AN on the desired user by projecting AN to the null space of all desired users' steering vectors from $K$ groups, which is written  as the form ${{\bf{H}}^H}\left( {{\theta _{dk}}} \right){{\bf{T}}_{AN}}{\bf{z}} = 0$ for all values of $k$ with~$k\in\left\{1,~2,~\cdots,~K\right\}$,  which can also be written as
\begin{equation}\label{K-Orth-Cond}
\left[{{{\bf{H}}^*}\left( {{\theta _{d1}}} \right)},\cdots,\\
{{{\bf{H}}^*}\left( {{\theta _{dK}}} \right)}
 \right]^T \cdot {{\bf{T}}_{AN}}{\bf{z}} = {{\bf{0}}_{\sum_{k=1}^KT_k \times 1}}.
\end{equation}
Let us define ${\bf{H}}^H_d=\big[{{\bf{H}}^H\left(\theta _{d1}\right)}^*,~\cdots,~{{{\bf{H}}^H}\left( {{\theta _{dK}}} \right)}^*$ $\big]^T
$, then the orthogonal condition in (\ref{K-Orth-Cond}) is reduced to
\begin{equation}\label{equation 29}
{\bf{H}}^H_d{{\bf{T}}_{AN}}{\bf{z}} = {{\bf{0}}_{\sum_{k=1}^KT_k \times 1}}\Rightarrow{{\bf{T}}_{AN}} = {{\bf{I}}_N} - {\bf{H}}_d{\left[ {{{{\bf{H}}}^H_d}{\bf{H}}_d} \right]^{ - 1}}{\bf{H}}^H_d.
\end{equation}
The added AN is intended  to interfere eavesdroppers, and the desired user groups receive no AN under ideal channel knowledge because we project AN onto the null space (NS) of all desired groups. To guarantee the existence of NS, the number of transmit antennas should be greater than the sum of receive antenna numbers of all desired groups of users.
\subsection{Proposed leakage-based method}
However, the Max-GRP plus NSP method does not exploit the effect of channel noise.  Now, we consider the  design of ${\bf{v}}_k$ and $\mathbf{T}_{AN}$  by using leakage idea with the help of variance of channel noise. Here, the  confidential-message power of group $k$ will be allowed to leak out towards other $K-1$ desired groups and eavesdropper one.  The corresponding precoding vector ${\bf{v}}_k$ is optimized  by the rule of maximizing the signal-to-leakage-and-noise ratio  (Max-SLNR)
\begin{align}\label{Max-SLNR}
\begin{array}{l}
\mathop {\max }\limits_{{{\bf{v}}_k}} ~~~\text{SLNR}\left( {{{\bf{v}}_k}} \right)\\
\text{subject~to}~{\bf{v}}_k^H{{\bf{v}}_k} = 1
\end{array}
\end{align}
where
\begin{align}
&\text{SLNR}\left( {{\mathbf{v}_k}} \right)=\left(\alpha _1^2\beta _1^2{P_s}tr\left\{ {{\bf{v}}_k^H{\bf{H}}\left( {{\theta _{dk}}} \right){{\bf{H}}^H}\left( {{\theta _{dk}}} \right){{\bf{v}}_k}} \right\}\right)\nonumber\\
&\Big[\text{tr}\Big(\alpha _1^2\beta _1^2{P_s}\sum\limits_{i = 1,i \ne k}^K {{\bf{v}}_k^H{\bf{H}}\left( {{\theta _{di}}} \right){{\bf{H}}^H}\left( {{\theta _{di}}} \right){{\bf{v}}_k}}  + \nonumber\\
&\alpha _1^2\beta _1^2{P_s}{\bf{v}}_k^H{\bf{H}}\left( {{\theta _e}} \right){{\bf{H}}^H}\left( {{\theta _e}} \right){{\bf{v}}_k} + \sigma _{dk}^2\Big)\Big]^{-1}.
\end{align}
According to the generalized Rayleigh-Ritz theorem, ${{\bf{v}}_k}\left( {k = 1,2, \cdots ,K} \right)$ is the eigenvector corresponding to the largest eigenvalue of  matrix
\begin{align}
&{\left[ {\sum\limits_{i = 1,i \ne k}^K {{\bf{H}}\left( {{\theta _{di}}} \right){{\bf{H}}^H}\left( {{\theta _{di}}} \right)}  + {\bf{H}}\left( {{\theta _e}} \right){{\bf{H}}^H}\left( {{\theta _e}} \right) + \frac{{\sigma _{dk}^2}}{{\alpha _1^2\beta _1^2{P_s}}}{{\bf{I}}_N}} \right]^{ - 1}}\cdot\nonumber\\
&~~~~~~~~~~~~~{\bf{H}}\left( {{\theta _{dk}}} \right){{\bf{H}}^H(\theta_{dk})}~~~~~~~~~~~~~~~~~~~~~~~
\end{align}
 Below, the AN  will be regarded as a useful signal to the eavesdropper group. By optimizing  the AN projection matrix ${\bf{T}}_{AN}$, the  leakage of AN power to $K$ desired groups  should be made as small as possible. This idea can be represented as
\begin{equation}
\begin{array}{l}
\mathop {\max }\limits_{{{\bf{T}}_{AN}}}~~~~~~~~\text{ANLNR}\left( {{{\bf{T}}_{AN}}} \right)\\
\text{subject~to} {\kern 1pt}~~\text{tr}\left( {{{\bf{T}}_{AN}}^H{{\bf{T}}_{AN}}} \right)=N-\sum_{k=1}^{K}T_k,
\end{array}
\end{equation}
which is called Max-ANLNR, where ANLNR stands for AN-leakage-and-noise-ratio defined by
\begin{small}
\begin{align}
&\text{ANLNR}\left( {{{\bf{T}}_{AN}}} \right)=\nonumber\\
&\frac{{tr\left\{ {{\bf{T}}_{AN}^H{\bf{H}}\left( {{\theta _e}} \right){{\bf{H}}^H}\left( {{\theta _e}} \right){{\bf{T}}_{AN}}} \right\}}}{{tr\left\{ {{\bf{T}}_{AN}^H\left( {\sum\limits_{i = 1}^K {{\bf{H}}\left( {{\theta _{di}}} \right){{\bf{H}}^H}\left( {{\theta _{di}}} \right)}  + \frac{{\sigma _e^2}}{{\alpha _2^2\beta _2^2{P_s}\left( {N -\sum_{i=1}^KT_k} \right)}}{{\bf{I}}_N}} \right){{\bf{T}}_{AN}}} \right\}}},
\end{align}
\end{small}
where, similar to Max-SLNR in (\ref{Max-SLNR}), all columns of AN projection matrix ${{\bf{T}}_{AN}}$ consists of the eigenvectors corresponding to the $N-
\sum_{k=1}^KT_k$ largest eigenvalues of the matrix given by
\begin{small}
\begin{align}
&{\left( {\sum\limits_{i = 1}^K {{\bf{H}}\left( {{\theta _{di}}} \right){{\bf{H}}^H}\left( {{\theta _{di}}} \right)} + \frac{{\sigma _e^2}}{{\alpha _2^2\beta _2^2{P_s}\left( {N -\sum_{k=1}^KT_k} \right)}}{{\bf{I}}_N}} \right)^{ - 1}} \cdot\nonumber\\
&~~~~~~~~~~~~~{\bf{H}}\left( {{\theta _e}} \right){{\bf{H}}^H}\left( {{\theta _e}} \right).~~~~~~~~~~~~~~~~~~~~~~~~
\end{align}
\end{small}
Therefore, we complete the design of the proposed Max-SLNR plus Max-ANLNR method.
\subsection{Complexity comparison and analysis}
In Table I,  we list the complexities of our proposed methods and conventional block diagonalization (BD)  method in \cite{choi}, respectively. {\color{blue}{BD is a classic precoding method for conventional MU-MIMO systems, and performs better than zero-forcing in terms of sum-rate  by allowing multi-antenna interference among antennas per user. Below, BD  is adopted as a performance benchmark.}} Here, $K$ stands for the number of desired groups and $T$ is the number of users per group with $T_1=\cdots=T_K=T$. From this table, it is seen that the complexities of BD and the proposed Max-GRP-NSP are the quadratic function of $K$ and the complexity of the Max-SLNR plus Max-ANLNR is a linear function of $K$ for fixed values of $M$, $N$ and $T$. The former increases quadratically with $K$ while the latter increases linearly with $K$. When $M$, $N$ and $K$ are fixed, the complexities of BD and the proposed Max-GRP-NSP are the cubic function of $T$ and the complexity of the Max-SLNR plus Max-ANLNR is a linear function. The former increases cubically with $T$ while the latter increases linearly with $T$. In general, $N\geq(KT+M)$, so the number of antennas $N$ at transmitter plays a dominant role in complexity. If we fix the values of $K$, $T$, and $M$, then the complexities of the three methods are cubic functions of $N$. In other words, they have the same magnitude complexity.

\section{simulations results}
Below, the performance of the two proposed methods will be evaluated. Simulation parameters are as follows:  $d=\lambda/2$, $N=16$, $K=2$, $T_1=T_2=M=2$,   $\theta_{d1}=\{30^{\circ}, 45^{\circ}\}$, $\theta_{d2}=\{120^{\circ}, 135^{\circ}\}$,  $\beta_1=\sqrt{0.9}$, {\color{blue}{and QPSK modulation is used to evaluate the BER performance.}}

\begin{figure}[h]
\centering
\subfigure[Group 1]{
\includegraphics[width=8.2cm]{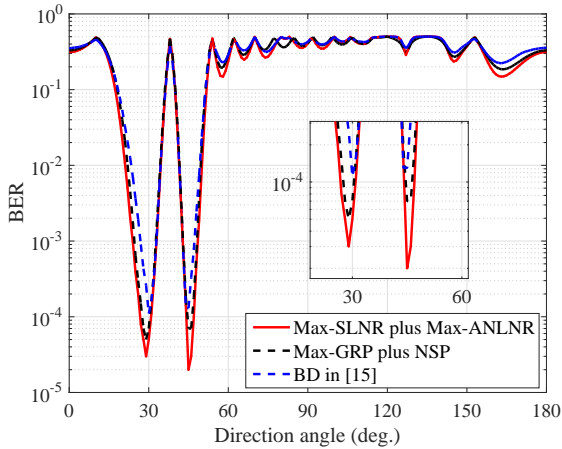}}
\subfigure[Group 2]{
\includegraphics[width=8.2cm]{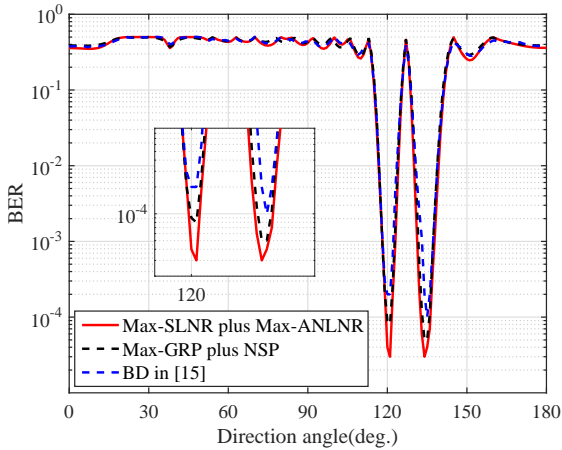}}
\caption{BER versus direction angle for two different desired user groups (SNR=14dB).}
\label{BER-DOA}
\end{figure}
Fig.~\ref{BER-DOA} shows the curves of BER versus direction angle of the two proposed methods for two desired groups, where the conventional BD method in \cite{choi}  is used as a performance reference. Parts (a) and (b) correspond to group 1 and group 2, respectively. It is seen from Part (a) that each BER curve  achieves  two local lowest values at the two desired directions of group 1: $30^{\circ}$ and $45^{\circ}$, and  rapidly rises up to more than 0.1. Clearly,  the BER performances of the proposed two methods  are better than that of BD along the desired directions. Additionally,  the  proposed two methods form two BER main beams around the two desired directions. Outside the two main beams, the BER performance degrades seriously. In other words, if eavesdroppers lie outside the two main beam, then it is extremely difficult for them to recovery the confidential message stream successfully. Observing Part (a), it is obvious that the Max-SLNR plus Max-ANLNR performs much better than the Max-GRP plus NSP at two desired directions. Part (b) plots the BER performance for group 2. Obviously, the same performance tendency is observed as Part (a).

\begin{figure}[h]
\centering
\subfigure[Group 1]{
\includegraphics[width=8.2cm]{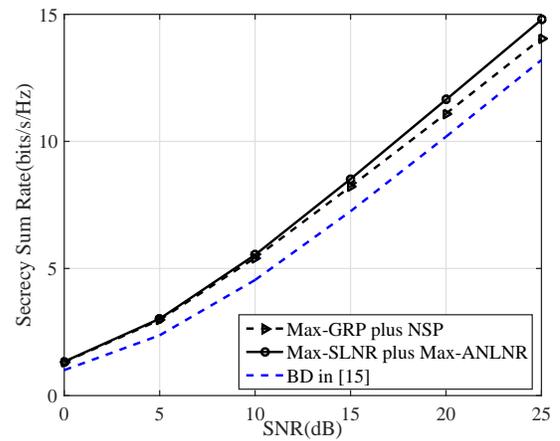}}
\subfigure[Group 2]{
\includegraphics[width=8.2cm]{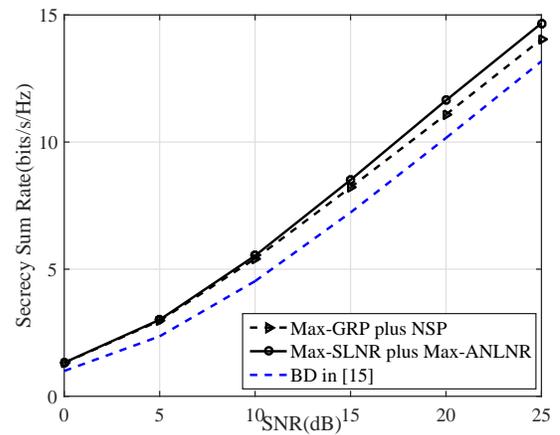}}
\caption{Group secrecy sum-rate versus {\color{blue}{SNR}} for two different desired user groups.}
\label{GSSR-DOA}
\end{figure}

Fig.~\ref{GSSR-DOA} illustrates the secrecy sum-rate (SSR) per group versus SNR of the two proposed methods for two desired groups. Here, Parts (a) and (b)  plot the SSR performance for groups 1 and 2, respectively. In the two parts, it is evident that the proposed two methods achieve an improvement over BD in SSR per group.  From Part (a), it follows  that the performance of the proposed Max-SLNR plus Max-ANLNR is slightly better than that of the Max-GRP plus NSP in the low SNR region, and much better than that of the latter in the medium and high SNR regions. As SNR increases, the performance gain over the Max-GRP plus NSP achieved by the proposed Max-SLNR plus Max-ANLNR increases gradually. Similar to Part (a), the same performance trend is seen in Part (b).
\begin{figure}[h]
\centering
\subfigure[Group 1]{
\includegraphics[width=8.2cm]{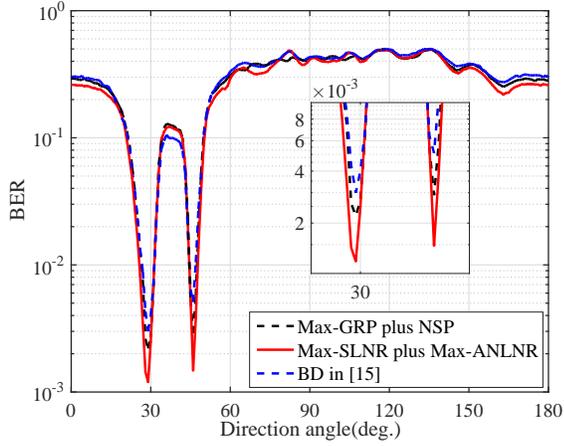}}
\subfigure[Group 2]{
\includegraphics[width=8.2cm]{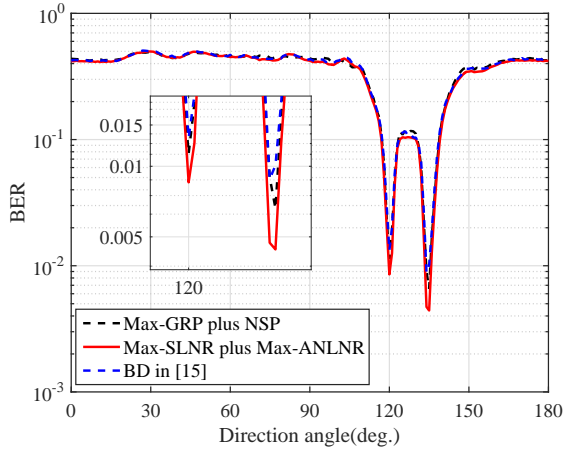}}
\caption{BER versus direction angle for two different desired user groups in the presence of direction measurement errors (SNR=14dB).}
\label{BER-DOA-Err}
\end{figure}
To evaluate the impact of direction measurement error on BER performance, Fig.~\ref{BER-DOA-Err} illustrates the BER performance of the two methods in the presence of direction angle measurement errors, where  direction angle measurement errors are approximately modeled and normalized as uniform distributed random variables in the interval $[-\frac{\Delta\theta_{max}}{BW},~\frac{\Delta\theta_{max}}{BW}]$ with  $\Delta\theta_{max}=5^\circ$ and $BW=2\lambda/(Nd)$, where $BW=2\lambda/(Nd)$ denotes the main beam bandwidth. From Parts (a) and (b), it is seen that the three methods become worse due to direction angle measurement errors. However, the proposed methods are more robust than BD.  Comparing Fig.~4 and Fig.~2, we find that there is an-order-of-magnitude performance loss on BER due to errors. Further, it is certain that the proposed Max-SLNR plus Max-ANLNR method performs still better along the desired directions than the Max-GRP plus NSP, and their BER performance degrades and shows approximately the same BER performance outside two main desired beams.

\section{Conclusion}
In this paper, two secure schemes, Max-GRP plus NSP and Max-SLNR plus Max-ANLNR, have been proposed  for  multi-cast DM scenario.
From simulation, we find  the proposed two methods behaves better than BD by means of BER and SSR per group.  The  Max-SLNR plus Max-ANLNR scheme performs much better than the Max-GRP plus NSP in accordance with BER and SSR per group regardless of direction measurement errors. Also, compared to the Max-SLNR plus Max-ANLNR, the Max-GRP plus NSP does not need variance of channel noise, and is simpler. Due to their good security and low-complexity, the proposed two schemes can be applied to the near future scenarios like unmanned aerial vehicle, satellite communications, and mmWave communications.

\ifCLASSOPTIONcaptionsoff
\newpage
\fi
\bibliographystyle{IEEEtran}
\bibliography{IEEEfull,REF}
\end{document}